\documentclass[10pt,conference]{IEEEtran}
\IEEEoverridecommandlockouts
\usepackage{cite}
\usepackage{amsmath,amssymb,amsfonts}
\usepackage{graphicx}
\usepackage{textcomp}
\usepackage{xcolor}
\def\BibTeX{{\rm B\kern-.05em{\sc i\kern-.025em b}\kern-.08em
    T\kern-.1667em\lower.7ex\hbox{E}\kern-.125emX}}
    
\usepackage{float}
\usepackage{amsmath,amssymb,amsfonts}
\usepackage[inline]{enumitem}     
\newcommand*\circled[1]{\tikz[baseline=(char.base)]{
            \node[shape=circle,fill,inner sep=0.5pt] (char) {\textcolor{white}{#1}};}}    

\usepackage{graphicx}
\usepackage{subcaption} 

\usepackage{textcomp}
\usepackage{multirow}  
\usepackage{xcolor}
\def\BibTeX{{\rm B\kern-.05em{\sc i\kern-.025em b}\kern-.08em
    T\kern-.1667em\lower.7ex\hbox{E}\kern-.125emX}}
    
\usepackage{hyperref} 
\usepackage{listings}

\usepackage{changepage}

\usepackage{ragged2e}
\usepackage{seqsplit}
\pdfmapfile{=rtxr.map} 

\usepackage[font=footnotesize,skip=1pt,labelfont=bf,tableposition=top]{caption}
\DeclareCaptionFont{white}{\color{white}}
\DeclareCaptionFormat{listing}{%
  \parbox{\textwidth}{\colorbox{gray}{\parbox{\textwidth}{#1#2#3}}\vskip-4pt}}
\hypersetup{colorlinks,linkcolor={blue},citecolor={blue},urlcolor={blue}}

\usepackage{algpseudocode}
\makeatletter
\def\BState{\State\hskip-\ALG@thistlm}
\makeatother
\usepackage{balance}
\usepackage[skins,listings,breakable]{tcolorbox}

\makeatletter
\let\old@lstKV@SwitchCases\lstKV@SwitchCases
\def\lstKV@SwitchCases#1#2#3{}
\makeatother

\usepackage{booktabs}
\usepackage{multirow}

\usepackage{lstlinebgrd}
\usepackage[utf8x]{inputenc}
\usepackage{ucs}
\usepackage{listings}
\usepackage{amsfonts}
\usepackage{amssymb}
\usepackage{tcolorbox}
\usepackage{colortbl}
\newtcbox{\highlight}[0]{boxsep=0pt,left=0pt,top=0pt,bottom=0pt,right=0pt,boxrule=0pt,arc=0pt,auto outer arc,colback=green,width=9cm}

\makeatletter
\let\lstKV@SwitchCases\old@lstKV@SwitchCases
\lst@Key{numbers}{none}{%
    \def\lst@PlaceNumber{\lst@linebgrd}%
    \lstKV@SwitchCases{#1}%
    {none:\\%
     left:\def\lst@PlaceNumber{\llap{\normalfont
                \lst@numberstyle{\thelstnumber}\kern\lst@numbersep}\lst@linebgrd}\\%
     right:\def\lst@PlaceNumber{\rlap{\normalfont
                \kern\linewidth \kern\lst@numbersep
                \lst@numberstyle{\thelstnumber}}\lst@linebgrd}%
    }{\PackageError{Listings}{Numbers #1 unknown}\@ehc}}
\makeatother

\lstdefinestyle{base}{
  mathescape,
  language=Java,
  basicstyle=\ttfamily\scriptsize,
  frame=lines,
  keywordstyle=\color{blue}\textbf,
  commentstyle=\color[rgb]{0.0,0.4,0.0}\scriptsize,
  extendedchars=true,         
  breaklines=true   
  showspaces=false,
  showstringspaces=false, 
   numbers=left,
    stepnumber=1,   
        tabsize=1,
    breaklines=true,      
    xleftmargin={0.5cm},
  moredelim=**[is][\color{green}]{!!}{!!},
  moredelim=**[is][\color{orange}]{^}{^},
  moredelim=**[is][\color{red}]{@}{@},
   breakindent=0pt,                
     }

\DeclareCaptionLabelFormat{andtable}{#1~#2  \&  \tablename~\thetable} 

\usepackage{diagbox}

\usepackage[linesnumbered,ruled,vlined]{algorithm2e}

\SetCommentSty{mycommfont}

\SetKwInput{KwInput}{Input}                
\SetKwInput{KwOutput}{Output}              

\SetAlgorithmName{Algo.}{}{}
\usepackage{soul}

\colorlet{soulred}{blue!10}
\DeclareRobustCommand{\hlcyan}[1]{{\sethlcolor{soulred}\hl{#1}}}

\usepackage{zlmtt}
\usepackage{fancyvrb,newverbs}
\definecolor{cverbbg}{gray}{0.93}

\newverbcommand{\cverb}
  {\setbox\verbbox\hbox\bgroup}
  {\egroup\colorbox{cverbbg}{\box\verbbox}}

\usepackage[square,sort,comma,numbers]{natbib}

\begin{document}

\title{\huge Leveraging Code Clones and Natural Language Processing for Log Statement Prediction
}

\author{\IEEEauthorblockN{Sina Gholamian}
\IEEEauthorblockA{\textit{University of Waterloo} \\
Waterloo, Canada \\
sgholamian@uwaterloo.ca}
}

\maketitle

\begin{abstract}
Software developers embed logging statements inside the source code as an imperative duty in modern software development as log files are necessary for tracking down runtime system issues and troubleshooting system management tasks.
Prior research has emphasized the importance of logging statements in the operation and debugging of software systems. 
However, the current logging process is mostly manual and \textit{ad hoc}, and thus, proper placement and content of logging statements remain as challenges. 
To overcome these challenges, methods that aim to automate log placement and log content, \textit{i.e.}, \textit{`where, what, and how to log'}, are of high interest. 
Thus, we propose to accomplish the goal of this research, that is \textit{``to predict the log statements by utilizing source code clones and natural language processing (NLP)''}, as these approaches provide additional context and advantage for log prediction. We pursue the following four research objectives: (RO1) investigate whether source code clones can be leveraged for log statement location prediction, (RO2) propose a clone-based approach for log statement prediction, (RO3) predict log statement's description with code-clone and NLP models, and (RO4) examine approaches to automatically predict additional details of the log statement, such as its verbosity level and variables.   
For this purpose, we perform an experimental analysis on seven open-source java projects, extract their method-level code clones, investigate their attributes, and utilize them for log location and description prediction.  
Our work demonstrates the effectiveness of log-aware clone detection for automated log location and description prediction and outperforms the prior work.\looseness=-1

\end{abstract}

\vspace*{-1mm}
\section{Introduction}
To gather feedback about computer systems' running state, it is a common practice for developers to insert logging statements inside the source code to have running programs' internal state and variables written to log files.
This logging process enables developers and system administrators to analyze log files for a variety of purposes~\cite{bertero2017experience}, such as anomaly and problem detection~\cite{xu2009detecting,fu2009execution}, log message clustering~\cite{makanju2009clustering,vaarandi2015logcluster}, system profile building~\cite{hassan2008industrial}, code quality assessment~\cite{shang2015studying}, and compression of log files~\cite{tang2011log,makanju2009clustering}. 
Additionally, the wealth of information in the logs has also generated significant industrial interest and thus has initiated the development of commercialized log processing platforms such as Splunk~\cite{urlsplunk} and Elastic Stack~\cite{urlelastic}. 

Due to the free-form text format of log statements and lack of a general guideline, adding proper logging statements to the source code remains a manual, inconsistent, and error-prone task~\cite{chen2017characterizing}. 
As such, methods to automate logging \textit{\textbf{location}} and predict the \textit{\textbf{details}}, \textit{i.e.}, the \textit{`static text'} and verbosity level of the logging statement, are well sought after. 
For example, the log print statement (LPS): {\small \hlcyan{log.warn(``Cannot find BPService for bpid=" + id)}}, contains a textual part indicating the context of the log, \textit{i.e.}, \textit{description}, \textit{``Cannot find BPService for bpid=''}, a \textit{variable} part, \textit{`id'}, and a log \textit{verbosity level},\textit{`warn'}, indicating the importance of the logging statement and how the level represents the state of the program~\cite{log4x}. 

For practical concerns such as I/O and development costs, the \textit{quantity}, \textit{location}, and \textit{description} of logging statements should be decided efficiently~\cite{jia2018smartlog}. 
Logging too little may result in missing important runtime information that can negatively impact the postmortem dependability analysis~\cite{yuan2012conservative}, and excessive logging can consume extra system resources at runtime and impair the system's performance as logging is an I/O intensive task~\cite{zhao2017log20,ding2015log2}. 
In addition, due to the current \textit{ad hoc} logging practices, developers often make mistakes in log statements or even forget to insert a log statement at all~\cite{hassani2018studying,li2021studying}. 
Therefore, prior studies have aimed to automate the logging process and predict whether a code snippet requires a logging statement by utilizing machine learning methods to \textit{train} a model on a set of logged code snippets, and then \textit{test} it on a new set of unlogged code snippets~\cite{fu2014developers,zhu2015learning} (supervised learning). 
A recent work~\cite{he2018characterizing} has shown similar code snippets are useful for log statement description (LSD) suggestions by evaluating their \textit{BLEU}~\cite{papineni2002bleu} and \textit{ROUGE}~\cite{lin2004rouge} scores, similar to \textit{Precision} and \textit{Recall}, respectively. 
Thus, in our research, we specifically seek to utilize source code clones for log statement prediction and suggestion. 

Our goal in this research is to utilize code clones as a paradigm to improve the log statement automation task. 
This will ensure consistency and a higher quality of logging compared to the current developers' \textit{ad hoc} logging efforts. 
To summarize, the objectives of this research are to first investigate the suitableness of source code clones for log statement prediction, uncover their shortcomings, and then leverage them for automated log location and description prediction based on selecting appropriate source code features~\cite{gholamian2020logging}. 
In addition, we utilize deep learning NLP approaches along with code clones to also predict the log statement's description. 
Through an empirical study of seven open-source software projects, we demonstrate the applicability of similar code snippets for log prediction, and further analysis suggests that log-aware clone detection can achieve high BLEU and ROUGE scores in predicting log statement's description.\looseness=-1

\section{Motivating Example}\label{casestudy}
Source code clones are exact or similar snippets of the code that exist among one or multiple source code projects~\cite{sajnani2016sourcerercc}.
There are four main classes of code clones~\cite{rattan2013software}: Type-1, which is simply copy-pasting a code snippet, Type-2 and Type-3, which are clones that show syntax differences to some extent, and finally Type 4, which represents two code snippets that are syntactically very different but semantically equal, \textit{e.g.}, iterative versus recursive implementations of \textit{Fibonacci} series in Figure~\ref{code_sample}. 
In this research, we focus on \textbf{method-level code clones} and call the tuple ($MD_{i}$, $MD_{j}$) a \textit{`clone pair'}. 
Figure~\ref{code_sample} shows that the logging pattern in the original code, $MD_i$ on Line 3 can be learned to suggest logging statements for its clone, $MD_j$, which is missing a logging statement. 
\begin{figure}[h]
\vspace*{-4mm}
\centering
\hspace*{-3mm}
\begin{minipage}{9cm}
\hspace*{-3mm}
\begin{minipage}{0.47\linewidth}
\scriptsize
\hspace*{-5mm}
\begin{lstlisting}[xleftmargin=-2cm,numbersep=2pt,linebackgroundcolor={%
    \ifnum\value{lstnumber}=1
            \color{blue!10}
    \fi
    \ifnum\value{lstnumber}=11
            \color{blue!10}
    \fi    
    },label={w_o_log_lvl_2},style=base]
//Original code - MD$_i$
int fibonacci(int n){
 log.info("Calculating Fibo sequence for %d.",n) 
 if(n==0||n==1)
  return n;
 else
  return fibonacci(n-1)+fibonacci(n-2);
}
\end{lstlisting}
\end{minipage}
\begin{minipage}{0.5\linewidth}
\scriptsize
\begin{lstlisting}[xleftmargin=-2cm,numbersep=2pt,linebackgroundcolor={%
    \ifnum\value{lstnumber}=1
            \color{blue!10}
    \fi
    \ifnum\value{lstnumber}=11
            \color{blue!10}
    \fi    
    },label={w_o_log_lvl_2},style=base]
//Clone Type 4 - MD$_j$
int getFibonacci(int n){
 if(n==0){return 0;}
 if(n==1){return 1;}
 int n_2th=0,n_1th=1,nth=1;
 for(int i=2;i<=n;i++){
  nth=n_2th+n_1th;
  n_2th=n_1th;
  n_1th=nth;}
 return nth;
\end{lstlisting}
\end{minipage}
\end{minipage}
\caption{Example for log prediction with code clones.}
\label{code_sample}
\vspace*{-2mm}
\end{figure}

\textbf{Practical Scenario.} To illustrate how our approach will be useful for developers during the development cycle of the software, we provide the following practical scenario. 
We consider a possible employment of our research as a recommender tool, which can be integrated as a plugin to code development environments, \textit{i.e.}, IDE software. 
Alex is a developer working on a large-scale software system and has previously developed method $MD_i$ in the code base. 
At a later time, Dave, Alex's colleague, is implementing $MD_{j}$. 
Our automated log suggestion\footnote{We use \textit{`suggestion'} and \textit{`prediction'} interchangeably.} approach can predict that if this new code snippet, $MD_{j}$, requires a logging statement by finding its clone, $MD_i$, in the code base. 
Then, the tool can suggest Dave, just in time, to add a log statement based on the prediction outcome.\looseness=-1 

\begin{figure*}[h]
\centering
\includegraphics[scale=.85]{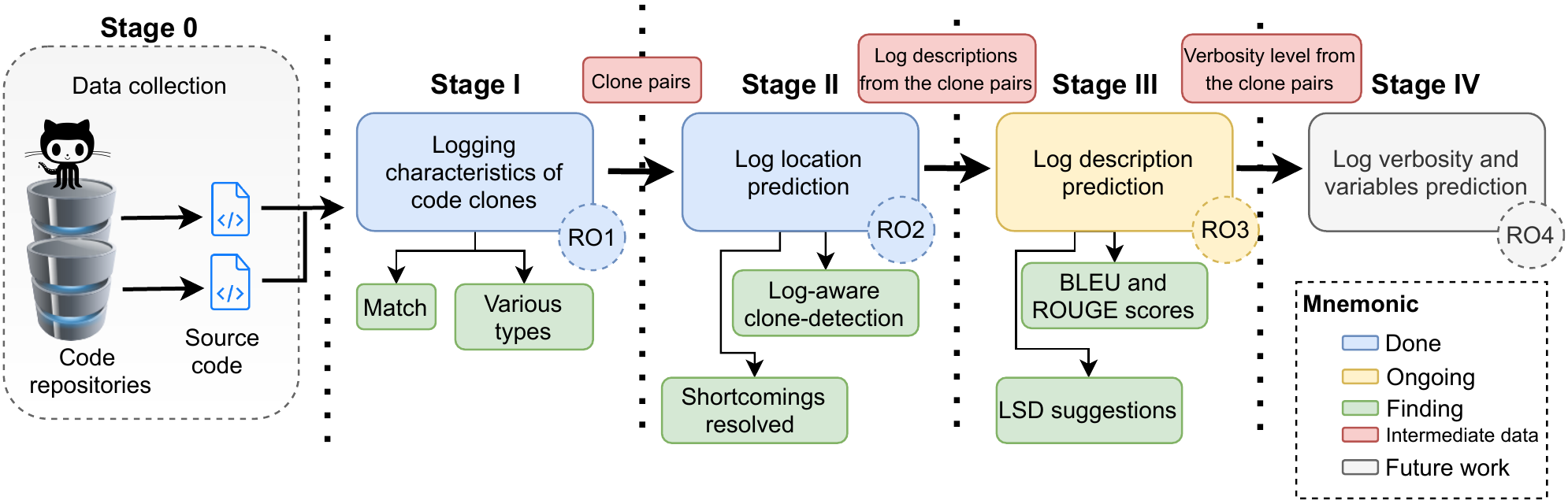}
\vspace*{-2mm}
\caption{Research steps, including objectives, intermediate data, and findings.}
\label{research_steps}
\vspace*{-5mm}
\end{figure*}

\section{Related Work}\label{rwork}
Prior work has tackled the automation of log statements with various approaches. 
Yuan \textit{et al.}~\cite{yuan2012conservative} proposed \textit{ErrorLog}, a tool to report error handling code, \textit{i.e.}, \textit{error logging}, such as \textit{catch clauses}, which are not logged and to improve the code quality and help with failure diagnosis by adding a log statement. Zhao \textit{et al.}~\cite{zhao2017log20} introduced \textit{Log20}, a performance-aware tool to inject new logging statements to the source code to disambiguate execution paths. 
\textit{Log20} introduces a logging mechanism that does not consider developers' logging habits or concerns. 
Moreover, it does not provide suggestions for \textit{logging descriptions}. 
Zhu \textit{et al.}~\cite{zhu2015learning} proposed \textit{LogAdvisor}, a learning-based framework, for automated logging prediction which aims to learn the frequently occurring logging practices automatically. 
Their method learns logging practices from existing code repositories for \textit{exception} and \textit{function return-value check} blocks by looking for textual and structural features within these code blocks with logging statements.    
Jia \textit{et al.}~\cite{jia2018smartlog} proposed an intention-aware log automation tool called \textit{SmartLog}, which uses an \textit{Intention Description Model} to explore the intention of existing logs, and Zhenhao et al.~\cite{li2020shall} categorized six block-level logging locations. 

Recently, Li \textit{et al}.~\cite{li2021studying} showed duplicate logging statements that are the outcome of shallow copy-pasting result in log-related anti-patterns (\textit{i.e.}, issues). 
Although their research has a negative connotation towards copy-pasted logging statements from code clones, it simultaneously shows the potential of code clones as a starting point for automated log suggestion and improvement. 
In other words, by automating and enhancing the log statements in the clone pairs, we can expedite the development process and avoid shallow copy-pasting that developers tend to do. 
Additionally, by automation, we reduce the risk of irregular and \textit{ad hoc} developers' logging practices, \textit{e.g.}, forgetting to log in the first place.

\section{Research Approach}
Based on the findings of He \textit{et al.}\cite{he2018characterizing} in logging description prediction based on \textit{edit distance}~\cite{ristad1998learning}, \ul{we hypothesize that similar code snippets, \textit{i.e.}, code clones, follow similar logging patterns which can be utilized for log statement location and description prediction}. 
Formally speaking, assuming set $CC_{MD_i}$ is the set of all code clones of Method Definition $MD_i$, if $MD_i$ has a log print statement (LPS), then its clones also have LPSs:\vspace{-1mm}
\begin{equation*} \label{eq1}
\vspace{-2mm}
\resizebox{0.91\hsize}{!}{
\hspace{-2mm}
$\exists LPS_i\in MD_i \implies \forall MD_{j}\in CC_{MD_i}, \exists LPS_j \in MD_{j}$ 
}
\vspace*{1mm}
\end{equation*} 

To evaluate the hypothesis, we guide our research with the following research objectives (\textbf{RO}s): 
\begin{itemize}
\item \textbf{RO1:} Demonstrate whether code clones are consistent in their logging statements.
\item \textbf{RO2:} Propose an approach to utilize code clones for log statement location prediction.
\item \textbf{RO3:} Provide logging description suggestions based on code clones and deep learning NLP models. 
\item \textbf{RO4:} Utilize clones for predicting other details of log statements such as log verbosity level and variables.  

\end{itemize}

Our research design comprises a preliminary data collection phase, Stage 0, and is followed by four stages, Stages I-IV, to address RO1-RO4, as illustrated in Figure~\ref{research_steps}. 
In the following, we provide the details of our methodology and current results for each RO.\looseness=-1   

\subsection{\textbf{RO1:} Demonstrate whether code clones are consistent in their logging statements and their log verbosity level.}
\textbf{\textit{Motivation}.} To enable code clones for log suggestion, we first require to compare their characteristics and show if clone pairs follow similar logging patterns.
\textbf{\textit{Approach}.} For this purpose, we select seven large-scale open-source Java projects, \textit{i.e.}, \textit{Apache Hadoop, Zookeeper, CloudStack, HBase, Hive, Camel, and ActiveMQ}, based on the prior logging research~\cite{chen2017characterizing,he2018characterizing}. 
These projects are well-logged, stable, and well-used in the software engineering community, and also they enable us to compare our results with prior work, accordingly. 
We extract methods with logging statements and then find their clones. 
\textbf{\textit{Evaluation}.} We evaluate the existence of log statements, their verbosity levels, and clone types.
\textbf{\textit{Results}.} The results show the majority of method clone pairs are consistent in their logging statements and their log verbosity levels also match to a high degree. 
Additionally, we find that the majority (in the range of 78\% to 90\%) of code clones are of Types 3 and 4, while the code pairs are matching in the existence of a logging statement. 
This observation signifies the effectiveness of code clones in suggesting the location of log statements in methods. 
In other words, although two snippets of clone pairs are syntactically different to a high degree, they still follow similar logging patterns.\looseness=-1
\vspace*{-1mm}
\subsection{\textbf{RO2:} Propose an approach to utilize code clones for log statement location prediction.}
\textbf{\textit{Motivation}.} Findings from RO1 show matching logging statements between clone pairs and motivate enablement of logging suggestions with code clones. 
The automated suggestion approach can help developers in making logging decisions and improve logging practices. 
\textbf{\textit{Approach}.} We initially observe and resolve two shortcomings of general-purpose clone detectors to make them more suitable for log prediction and reduce false positive and false negative cases~\cite{gholamian2021underreview}. 
We then utilize the clone pairs for suggesting logging statements for the methods which are missing an LPS by finding their clone pairs with a logging statement (Stage II in Figure~\ref{research_steps}). 
\textbf{\textit{Evaluation}.} We evaluate the performance of our approach by measuring \textit{Precision}, \textit{Recall}, \textit{F-Measure}, and \textit{Balanced Accuracy} (BA) on the set of the seven selected projects.
\textbf{\textit{Results}.} Considering the average of BA values, our log-aware clone detection approach, LACCP, brings 15.60\% improvement over Oreo~\cite{saini2018oreo} across the experimented projects. 
With the higher accuracy that LACCP brings, it enables us to provide more accurate clone-based log statement suggestions.

\subsection{\textbf{RO3:} Provide logging description suggestions based on code clones and NLP models.}
\textbf{\textit{Motivation}.} Based on the experiment results for predicting the location of logging statements in RO2 and the additional available context from the clone pairs, \textit{i.e.}, the logging statement description available from the original method, $MD_i$, we notice it is a valuable research effort to explore whether it is also possible to predict the logging statements' \textit{description} automatically. 
With satisfactory performance, an automated tool that can predict the description of logging statements will be a great aid, as it can expedite the software development process and improve the quality of logging descriptions. 
\textbf{\textit{Approach}.} We base our method on the assumption that clone pairs tend to have similar logging statement descriptions. 
This assumption comes from the observations in predicting log statements for clone pairs. 
As logging descriptions explain the source code surrounding them, it is intuitive for similar code snippets to have comparable logging descriptions. 
Based on this assumption, we propose a deep learning-based method that combines code clones with NLP learning approaches (NLP CC'd). 
In particular, to generate the LSD for a logging statement in $MD_j$, we extract its corresponding code snippet and leverage LACCP to locate its clone pairs. 
Laterally, the NLP model provides next word suggestions for the LSDs from the knowledge base available in the training set for each project. 
\textbf{\textit{Evaluation}.} To measure the accuracy of our method in suggesting the log description, we utilize BLEU~\cite{papineni2002bleu} and ROUGE~\cite{lin2004rouge} scores. 
These scores are well-established for validating the usefulness of an auto-generated text in prior software engineering and machine learning research, such as comment and code suggestion~\cite{allamanis2018survey} and description prediction~\cite{he2018characterizing}.
\textbf{\textit{Results}.} We experiment on seven open-source Java systems, and our analysis shows that by utilizing log-aware clone detection and NLP, our hybrid model, (\textit{NLP CC'd}), achieves 40.86\% higher performance on BLEU and ROUGE scores for predicting LSDs when compared to the prior research~\cite{he2018characterizing}, and achieves 6.41\% improvement over the No-NLP version~\cite{gholamian2021underreview}.

\subsection{\textbf{RO4:} Utilize code clones for predicting other details of log statements such as log verbosity level and variables.}
\textbf{\textit{Motivation}.} Besides log statement location and its LSD, prediction of other details of log statements such as \textit{Log verbosity level (LVL)} and its \textit{variables (VAR)} are useful research efforts and the focus of prior research~\cite{li2017log,anu2019approach,li2021deeplv}, as they further help the developers in more systematic logging and resolve suboptimal choices of log levels and variables~\cite{li2021deeplv}. 
\textbf{\textit{Approach}.} Log-aware clone detection, LACCP, is reasonably extendable to predict LVL and VAR alongside the LSD suggestion. 
Since we have access to the source code of the method that we are predicting the logging statement for and its clone pair code snippet, a reasonable starting point is to suggest the same LVL as of its clone pair, and then augment it with additional learning approaches such as~\cite{li2017log,anu2019approach} for more sophisticated LVL prediction. 
For VAR prediction, our approach can be augmented with deep learning~\cite{liu2019variables} and static analysis of the code snippet under consideration~\cite{yuan2012improving} to include log variables suggestions along with the predicted LSD. 
\textbf{\textit{Preliminary evaluation and results}.} Our preliminary analysis for the evaluated projects shows that code clones match in their verbosity levels in the range of (92, 97)\%, which confirms that using the verbosity level of the clone pair, $MD_i$, is a good starting point for log verbosity level suggestions for $MD_j$. 
We are pursuing RO4 as our future work and will provide additional results and findings subsequently.

\section{Discussion}
In this section, we compare and discuss the significance of our approach in relation to other existing log prediction and suggestion techniques.

\textbf{Method-level log prediction rationale.} Although clone detection (and subsequently, log statement prediction) can be performed in different granularity levels, such as files, classes, methods, and code blocks, however, method-level clones appear to be the most favorable points of re-factoring for all clone types~\cite{kodhai2014method}. 
We emphasize that our approach also includes all of the logging statements which are nested inside more preliminary code blocks within method definitions, \textit{viz.}, logging statements nested inside code blocks, such as \textit{if-else} and \textit{try-catch}.\looseness=-1 

\textbf{Comparison.}
Orthogonal to our research, prior efforts such as \cite{zhu2015learning}, \cite{jia2018smartlog}, and~\cite{li2020shall} have proposed learning approaches for logging statements' \textit{location} prediction, \textit{i.e.}, \textit{where to log}. 
The approaches in \cite{zhu2015learning}, \cite{jia2018smartlog} are focused on error logging statements (ELS), \textit{e.g.}, log statements in \textit{catch clauses}, and are implemented and evaluated on C\# projects. 
Li \textit{et al.}~\cite{li2020shall} provide log location suggestions by classifying the logged locations into six code-block categories. 
Different from these works, our approach does not distinguish between error and normal logging statements, is evaluated on open-source Java projects, and leverages logging statement suggestions at method-level by observing logging patterns in similar code snippets, \textit{i.e.}, clone pairs.
 
\textbf{Significance.} Prior approaches~\cite{zhu2015learning,li2020shall} rely on extracting features and training a learning model on logged and unlogged code snippets. 
Thus, they can predict if a new unlogged code snippet needs a logging statement by mapping its features to the learned ones. 
Although these methods initially appear similar to our approach in extracting log-aware features from code snippets~\cite{gholamian2020logging}, we believe our approach has an edge over the prior work. 
Because we also have access to the clone pair of the code under development, \textit{i.e.}, $MD_i$ in $(MD_i, MD_{j})$, this enables us to obtain and leverage the additional data from $MD_i$ to predict other aspects of log statements, \textit{e.g.}, LSD, which the prior work is unable to do. 
The significance of our approach becomes apparent in LSD prediction as we utilize the LSD of the clone pair as a starting point for suggesting the LSD of the new code snippet. 
Thus, our approach not only complements the prior work in providing logging suggestions for developers as they develop new code snippets, but it also has an edge over them by providing additional context for further prediction of LPS details, such as the LSD and the log's verbosity level. 

\section{Summary of Contributions}
The contributions that become available as the outcomes of our research are as follows: 
\begin{enumerate*}[label=\protect\circled{\arabic*}]
\item In RO1, we perform an experimental study on logging characteristics of code clones and show the potential for utilizing clone pairs for logging suggestions.
\item In RO2, we introduce a log-aware clone detection tool (\textit{LACCP})~\cite{gholamian2020logging} for log statements' \textit{`location'} prediction, and resolve two clone detection shortcomings for log prediction and provide experimentation on seven projects and compare it with general-purpose state-of-the-art clone detector, Oreo~\cite{saini2018oreo}.
\item In RO3, we initially show the natural characteristics of software logs and that enables us to utilize our findings for the application of NLP for LSD prediction~\cite{gholamian2021naturalness}. 
We then propose a deep-learning NLP approach, \textit{NLP CC'd}, to work in collaboration with \textit{LACCP} to automatically suggest log statements' descriptions.  
We calculate the BLEU and ROUGE scores for our auto-generated log statements' \textit{descriptions} by considering different sequences of LSD tokens, and compare our performance with the prior work~\cite{he2018characterizing}. 
\item Finally, as future work in RO4, we investigate the log verbosity level and variables prediction based on the information available through code clone pairs. 
\end{enumerate*}

Thus far, our research findings have been published for RO1 and RO2 in \textit{ACM Symposium on Applied Computing (ACM SAC)}~\cite{gholamian2020logging} and \textit{IEEE/ACM Conference on Mining Software Repositories (MSR)}~\cite{gholamian2021naturalness}, respectively. 
We have also evaluated the trade-offs associated with the cost of logging statements in our paper accepted in the \textit{International Symposium on Reliable Distributed Systems (SRDS)}~\cite{gholamian2021distributed}. 
Lastly, the research paper summarizing our contributions for RO3 is currently under review~\cite{gholamian2021underreview}.

\section{Conclusions and Future Work}
The process of software logging is currently manual and lacks a unified guideline for choosing the location and content of log statements. 
In this research, with the goal of enhancing log statement automation, we present a study on the location and description of logging statements in open-source Java projects by applying code clones and deep-learning NLP models. 
We compare the performance of our proposed approaches, LACCP and NLP CC'd, for log location and description prediction, and show their superior performance compared to prior work. 
As our future work in RO4, we will provide automated suggestions for other details of the LPS, such as its verbosity level and variables.

\bibliographystyle{IEEEtran}
\bibliography{ASE_2021}

\end{document}